\documentclass[a4paper,fleqn,usenatbib]{mnras}

\usepackage{mathptmx}
\usepackage[T1]{fontenc}
\usepackage{ae,aecompl}

\usepackage{graphicx}   % Including figure files
\usepackage{amsmath}    % Advanced maths commands
\usepackage{amssymb}    % Extra maths symbols
\usepackage{lipsum}
\usepackage{mathtools}
\usepackage{cuted}

\newcommand{\Msun}{\ensuremath{{\rm M}_{\odot}}}

\newcommand{\avg}[1]{\ensuremath{\left\langle \,#1\, \right\rangle}}

\newcommand{\der}{\ensuremath{{\rm d}}}

\newcommand{\be}{\begin{equation}}
\newcommand{\ee}{\end{equation}}
\newcommand{\bear}{\begin{eqnarray}}
\newcommand{\ear}{\end{eqnarray}}
\newcommand{\nline}{\nonumber \\}
\newcommand{\f}{\frac}

\title[Unbiased reionization constraints]{Unbiased constraints on reionization model parameters in presence of the foreground wedge}

\author[Raut \& Choudhury]{Dinesh Raut$^{1}$\thanks{Email:dinesh@ncra.tifr.res.in} and Tirthankar Roy Choudhury$^{1}$ \\
$^{1}$National Centre for Radio Astrophysics, TIFR, Post Bag 3, Ganeshkhind, Pune 411007, India
}

\pubyear{2017}

\begin{document}
\label{firstpage}
\pagerange{\pageref{firstpage}--\pageref{lastpage}}
\maketitle

\begin{abstract}
A possible way to disentangle the redshifted 21~cm signal of neutral hydrogen (HI) in the Epoch of Reionization (EoR) from the much larger astrophysical foregrounds is to restrict the analysis to a wedge-shaped region in the Fourier space. The foregrounds are confined to this wedge because of their smooth spectral properties which allows one to estimate the HI power spectrum in the foreground-free portion, known as the reionization window. The estimate of the spherically averaged power spectrum in the window, however, differs from the true value because of the line of sight anisotropies in the signal. The difference can be estimated and taken into account for a given reionization model by expanding the power spectrum in terms of the Legendre polynomials. This corrected power spectrum, called as the clustering wedges, can then be used for comparing the model predictions with the observations. In this work, we carefully examine whether the clustering wedges are appropriate for such model comparisons and whether they provide truly unbiased constraints on the EoR parameters. We use the excursion-set based semi-numerical simulations, coupled with MCMC-based statistical methods, for our analysis. We find that the clustering wedges not only yield faithful constraints, the statistical uncertainties on the parameter values too are comparable to those obtained using the data without the presence of any foregrounds. We also find that the clustering wedges are helpful in breaking the degeneracies between different EoR model parameters.
\end{abstract}


\section{Introduction}

One of the most promising ways of studying the universe during the epoch of reionization is through the redshifted 21~cm line originating from the neutral hydrogen (HI) in the intergalactic medium (IGM) \citep[for reviews, see][]{2006ARA&A..44..415F,2006PhR...433..181F,2009CSci...97..841C,2012RPPh...75h6901P}. The main challenge in observing this cosmological signal lies in differentiating it from the otherwise much larger astrophysical foregrounds \citep[see, e.g.,][]{2006PhR...433..181F,2010MNRAS.409.1647J} which are estimated to dominate the HI signal by a factor of  $10^4\mbox{-}10^5$ \citep{2002ApJ...564..576D,2003MNRAS.346..871O,2004MNRAS.355.1053D,2008MNRAS.385.2166A}. 

One expects the detection of the cosmological signal in presence of such a large foreground component is because the foregrounds are expected to be smooth functions of frequency while the signal is expected to fluctuate rapidly. One can thus subtract out the foregrounds by carefully modelling their frequency dependence \citep{2005ApJ...625..575S,2006ApJ...638...20B,2006ApJ...650..529W,2008MNRAS.391..383G,2009MNRAS.398..401L,2009MNRAS.394.1575L,2009MNRAS.397.1138H,2010MNRAS.405.2492H,2011PhRvD..83j3006L,2011MNRAS.413.2103P,2012MNRAS.423.2518C,2015MNRAS.447.1973B,2015MNRAS.452.1587G}  or by their isolation in the Fourier domain \citep{2010ApJ...724..526D,2012ApJ...745..176V,2012ApJ...752..137M,2012ApJ...757..101T,2012ApJ...756..165P,2013ApJ...768L..36P,2013ApJ...770..156H,2014PhRvD..90b3018L,2014PhRvD..90b3019L,2015ApJ...804...14T}. The foregrounds, being smooth, would not occupy all of Fourier or $k$-space and hence certain regions of $k$-space can be made foreground-free by a careful choice of the window function. The region where foregrounds are expected to be dominant is expected to be of wedge shape in $k_{\perp} \mbox{-} k_{\parallel}$ space, where $k_{\parallel}$ and $k_{\perp}$ are magnitudes of the component of the Fourier modes in the direction parallel and perpendicular to the line of sight (LOS). The remaining portion of the $k$-space, where the fluctuating 21~cm signal is expected to be dominant and hence can be detected more easily, is called as the EoR window. This method of foreground avoidance is already being used for estimation of EoR signal by experiments like PAPER \footnote{http://eor.berkeley.edu/} \citep{2015ApJ...809...61A} and MWA\footnote{http://www.mwatelescope.org/} \citep{2016ApJ...833..213P}. These experiments are expected to give an estimate of the statistical fluctuations of the 21~cm signal on various scales in terms of the power spectrum of 21~cm radiation. 

Being limited to the EoR window, these predictions will not be able to sample the whole $k$-space and so the power spectrum measured need not be a complete representation of the underlying cosmological signal. Had the signal being isotropic, this distinction would have led to only an increase in errors and no bias. But due to the presence of peculiar velocities of HI clouds, the cosmological signal is not isotropic and so its measurement would, in general, depend on the region of $k$-space sampled (which in our case corresponds to the EoR window). The difference, established in earlier studies \citep{2014PhRvD..90b3018L,2016MNRAS.456...66J,2018MNRAS.475..438R} can be estimated and accounted for \citep{2013MNRAS.435...64K,2018MNRAS.475..438R}, at least for large scales (where upcoming experiments would be focussing upon),  by exploiting the smooth nature of the mean power spectrum.  In our previous paper \citep{2018MNRAS.475..438R}, we introduced a procedure which involves computing the moments of the complete $k$-space power spectrum in the basis of Legendre polynomials and consequently estimate the wedge bias using these moments. The incomplete $k$-space moments, which are computed using only the modes inside the EoR window, are called the clustering wedges and are natural candidates for comparing observations with theoretical predictions. 

As a follow up of our previous work, in this paper, we attempt to answer the question whether these wedges can indeed facilitate a faithful comparison of theory with observations. In order to do so, we make use of the semi-numerical simulations of reionization to generate the underlying signal. The main purpose of this work is to examine whether usage of the clustering wedges can enable faithful recovery of the underlying model parameters with the upcoming telescopes like the SKA. We use the Monte Carlo Markov Chain (MCMC) based statistical methods to carry out the analysis which has been previously shown to be useful while comparing semi-numerical simulations with data \citep{2015MNRAS.449.4246G}. The paper is organized as follows: The simulations of the 21~cm signal are discussed in Section 2. In Section 3, we give a brief account of clustering wedges. In Section 4, we discuss our methodology of the model selection procedure using the Markov Chain Monte Carlo (MCMC) sampling algorithm. In Section 5, we present the results obtained using  the method and compare it to the idealized foreground free scenario. We also discuss the origin of degeneracy in the parameter predictions for complete and partial $k$-space coverages. In the Section 6, we try to estimate the parameter constraints arising from the foreground modelling approach. Finally in Section 7, we summarize and discuss our results. The cosmological parameters used for this study are $\Omega_m = 0.308$, $\Omega_{\Lambda} = 1 - \Omega_m$, $h = 0.678$, $\Omega_b = 0.049$, $\sigma_8 = 0.84$ and $n_s = 0.966$ \citep{2016A&A...594A..13P}.

\section{Simulations of the 21~cm signal}
In this section, we discuss the methodology used for the generation of maps of 21~cm brightness temperature distribution. The method uses two important theoretical ideas, both based on excursion set theory. The first one yields a halo catalogue from an overdensity field obtained from a low-resolution $N$-body simulation, while the second one generates an ionization field using the same overdensity field and the halo catalogue generated. The main steps can be summarized as follows:

\begin{itemize}
\item We first run a dark matter-only $N$-body simulation for generating the density and velocity fields. The simulations are performed using publicly available code {\sc gadget-2}\footnote{\tt https://wwwmpa.mpa-garching.mpg.de/gadget/} \citep{2005MNRAS.364.1105S}. We have used a cubical box of length $L_{box} = 200h^{-1} {\mathrm {cMpc}}$ having $512^3$ particles, thus giving a particle mass of $\approx 5.05 h^{-1}\times10^9 {\Msun}$  .

\item Since the particle mass is larger than that of typical haloes which are expected to host the first stars, it is not possible to identify the haloes using any group-finder algorithm. Instead, we use the density field on a uniform grid and we apply the conditional mass function of \citet{2002MNRAS.329...61S} to populate the grid cells with haloes having mass larger than a specific value $M_{\rm min}$. Our method is based on the one described in \citet{2016JCAP...03..001S}, except that we tune the parameters of the ellipsoidal collapse barrier of \citet{2002MNRAS.329...61S} so as to match the halo mass function fitted to simulations \citep{2001MNRAS.321..372J}. The halo mass function in each cell is then translated into a collapse mass fraction $f_{\rm{coll}}$, which would be a function of the parameter $M_{\rm {min}}$. This function is calculated and stored for each grid cell separately. 

\item Once we have the density field and $f_{\rm{coll}}$ for each cell, the ionization field can then be generated using a semi-numerical algorithm of \citet{2007ApJ...654...12Z,2009MNRAS.394..960C} which requires $N_{\rm {ion}}$ as an additional parameter. This parameter can be interpreted as the number of ionizing photons in the IGM per collapsed baryon in haloes. The 21~cm brightness temperature is then obtained using the expression \citep{1997ApJ...475..429M,2006PhR...433..181F}
\be
\delta T_b({\mathbf x}) = 27 x_{\rm{HI}}({\mathbf x}) [1+\delta_b({\mathbf x})] \left( \f{\Omega_b h^2}{0.023} \right)
\times \left( \f{0.15}{\Omega_m h^2} \f{1+z}{10} \right)^{1/2} {\rm {mK}}
\label{eq:dTb}
\ee 
This equation assumes that the $T_{\rm{spin}} \gg T_{\rm{CMB}}$. The semi-numerical scheme is very efficient in generating the brightness temperature distribution and hence is much suited for doing the MCMC model selection \citep{2015MNRAS.449.4246G}.
 
\item The effects of peculiar velocities were then incorporated using the MM-RRM scheme of  \citet{2012MNRAS.422..926M}.
\end{itemize}

\begin{figure}
\centering
\includegraphics[width=0.35\textwidth]{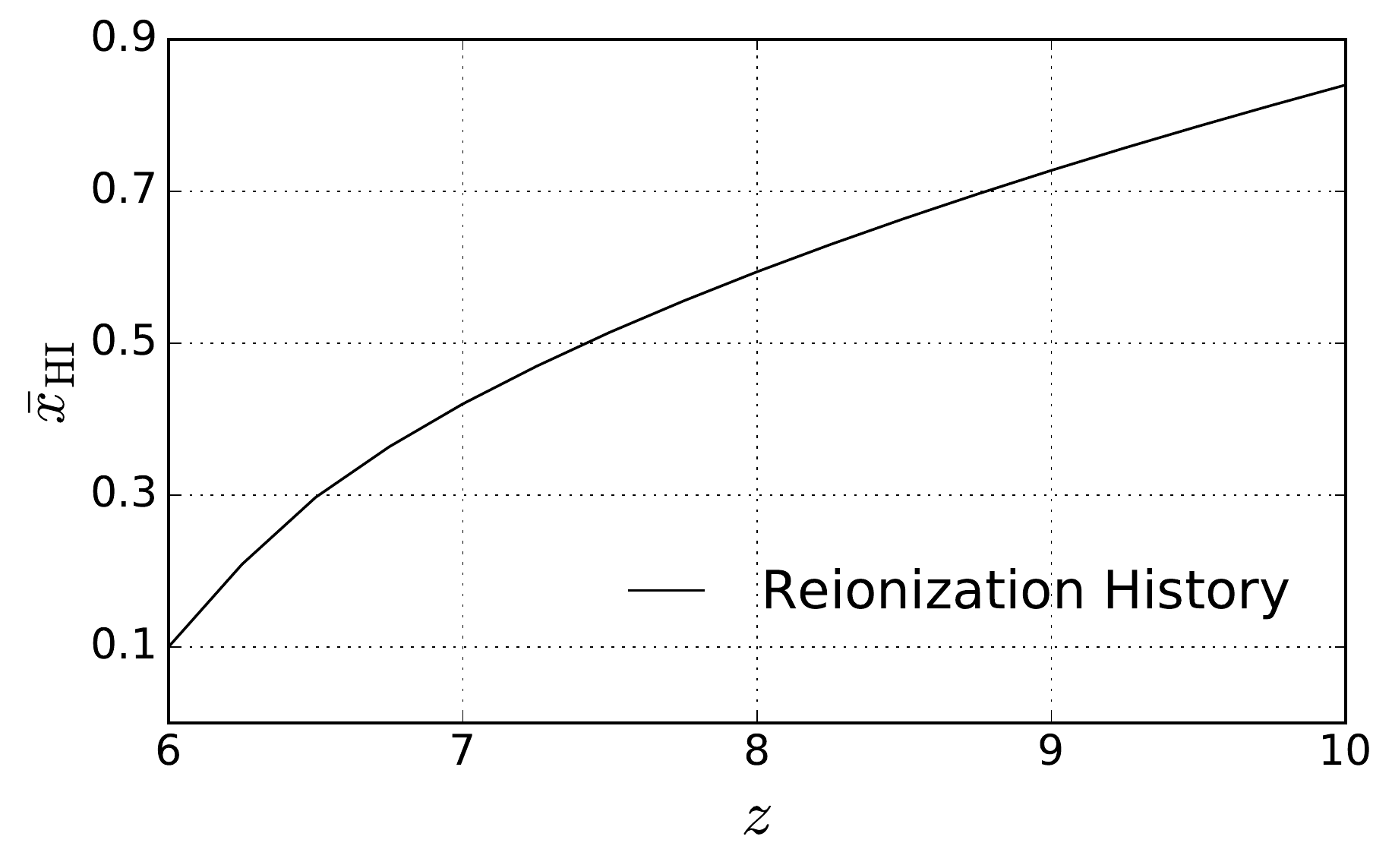}
\caption{The dependence of mean neutral hydrogen fraction on redshift used in this study.}
\label{fig:hist}
\end{figure}

The temperature maps, which now also include the effects of redshift space distortion (RSD), can then be used for further analysis. The RSD makes the observed cosmological signal anisotropic, thus the value of power spectrum computed within the EoR window will, in general, differ from its true estimate. We choose a fiducial reionization history which is equivalent to the ``Very Late'' history of \citet{2016MNRAS.463.2583K} and thus is consistent with the CMBR and Ly$\alpha$ absorption data. The evolution of $\bar{x}_{\rm HI}(z)$ for this history is shown in Figure~\ref{fig:hist}. The fiducial values of the reionization parameters are $M_{\rm{Min}} = 10^8 M_{\odot}$ (i.e., only atomic cooling is assumed to be efficient in haloes) and $N_{\rm{ion}}$ chosen separately for each redshift such that it reproduce the desired $\bar{x}_{\rm{HI}}(z)$ as in Figure~\ref{fig:hist}.

\section{Clustering wedges for Model comparison}

As discussed in our earlier paper \citep{2018MNRAS.475..438R}, the power spectrum measured in the EoR window can be estimated accurately using the formalism of clustering wedges. In this section, we briefly review the formalism. 

The brightness temperature in the Fourier space is given by,
\be
\avg{\delta \hat{T}_b(\mathbf{k}) ~\delta \hat{T}^*_b(\mathbf{k'})} = (2 \pi)^3~\delta_D(\mathbf{k} - \mathbf{k'})~P(\mathbf{k}),
\ee
where $\delta \hat{T}_b(\mathbf{k})$ is the Fourier transform of $\delta \hat{T}_b(\mathbf{x})$. We will be using the dimensionless power spectrum defined to be
\be
\Delta^2(\mathbf{k}) = \f{k^3 P(\mathbf{k})}{2 \pi^2}.
\ee
The spherically averaged power spectrum is obtained by averaging $\Delta^2(\mathbf{k})$ over all possible angles
\be
\Delta^2_0(k) = \f{1}{2} \int_{-1}^1 \der \mu~\Delta^2(\mathbf{k}) = \int_0^1 \der \mu~\Delta^2(\mathbf{k}),
\ee
where $\mu=k_{\parallel}/k$ corresponds to the cosine of the angle that wavevector ${{\mathbf k}}$ makes with the LOS direction. The other $k$-component, being perpendicular to the LOS direction, is denoted as $k_{\perp}$. The equation of line separating the Foreground dominated region from the EoR window is given by,
\be
k_{\parallel} \leq C~k_{\perp},~~
C = \sin \theta_{\rm FoV}~\f{D_c(z) H(z)}{c (1+z)},
\label{eq:C}
\ee
where $D_c(z)$ is comoving distance to redshift $z$, $H(z)$ is the Hubble constant at redshift $z$ and $\theta_{\rm {FoV}}$ the field of view in radians. In terms of variable $\mu$, the line is defined to be the one separating the two regions $1 \ge \mu \ge \mu_{\rm{min}}$ and $-1 \le \mu \le -\mu_{\rm{min}}$ where 
\be
\mu_{\rm min} = \f{C}{\sqrt{1 + C^2}}.
\label{eq:mumin}
\ee
The power spectrum measured in the window is thus
\be
\Delta^2_{0, {\rm win}}(k) = \f{1}{1 - \mu_{\rm min}} \int_{\mu_{\rm min}}^1 \der \mu~\Delta^2(\mathbf{k}).
\ee

The actual value of $\mu_{\rm min}$ in an experiment will depend on various factors like the beam pattern of the interferometric elements and also the accuracy of the calibrating the same. In absence of these details for the SKA1-Low, we choose to work with a rather conservative value $\mu_{\rm min} = 0.9$. This value of $\mu_{\rm{min}}$ corresponds to a field of view very close to the horizon wedge $\theta_{\rm{FoV}}=90^o$ \citep{2016MNRAS.456...66J} and hence would be valid even when there is significant foreground contamination from the side-lobes of the beam.

\begin{figure}
\includegraphics[width=0.45\textwidth]{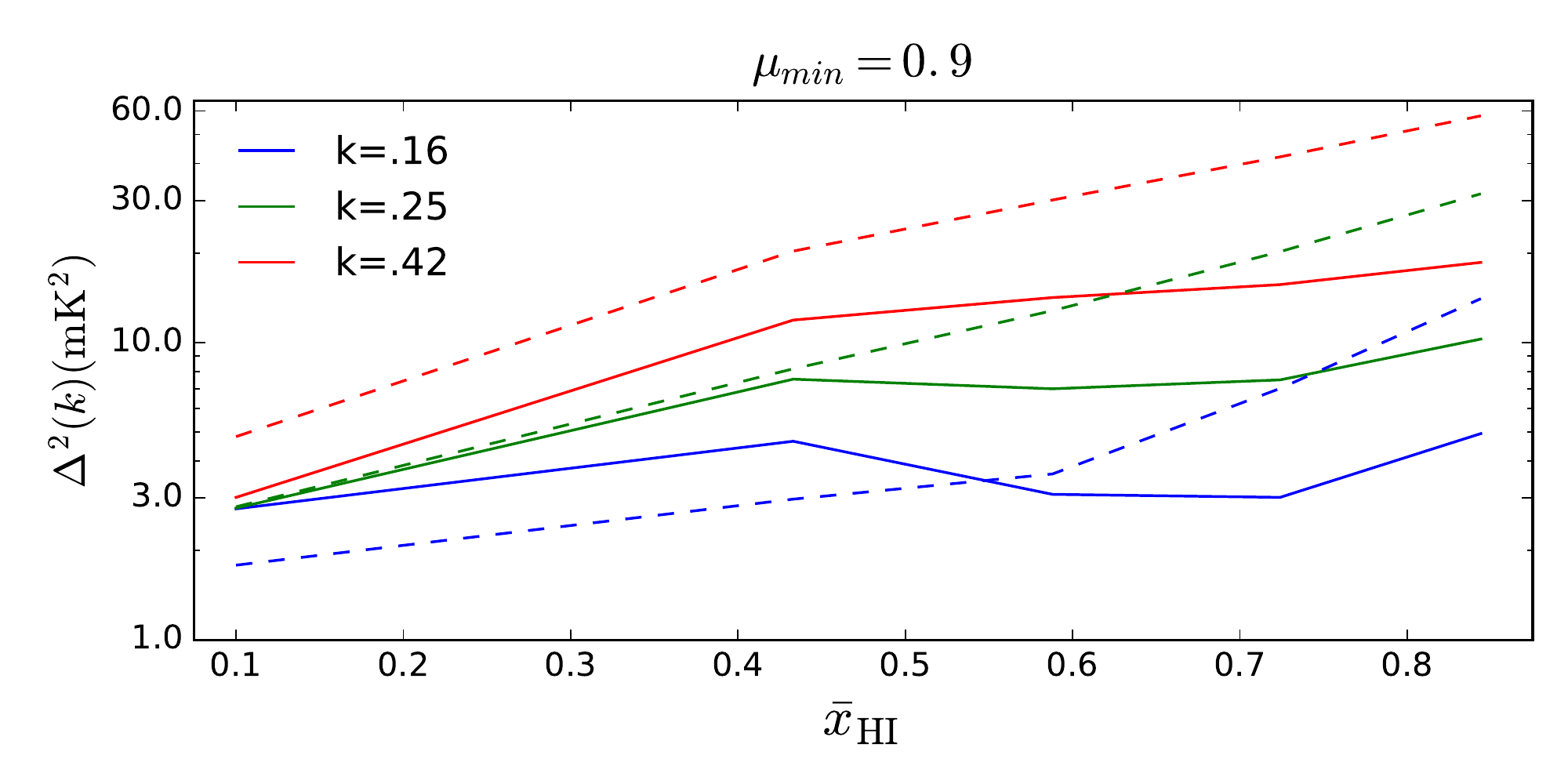}
\caption{Dependence of power spectrum on the extent of wedge for 3 typical modes as a function of mean neutral hydrogen fraction. The solid lines correspond to the entire $k$-space measurements while the dashed lines correspond to the measurements confined to the EoR window with $\mu_{\rm{min}}=0.9$.} 
\label{fig:jensen}
\end{figure}

We show the evolution of the average power spectra for our fiducial reionization history, computed over the window and over the entire $k$-space, in Figure~\ref{fig:jensen}. As seen from the figure, the average power spectrum computed over the window differs from the average of power spectrum computed using the entire $k$-space. One can now use the earlier formalism to compute the power spectrum in the window to be
\bear
\Delta^2_{0, {\rm win}}(k) &=& \sum_{l~{\rm even}} \Delta^2_l(k) \left[\f{1}{1 - \mu_{\rm min}} \int_{\mu_{\rm min}}^1 \der \mu~{\cal P}_l(\mu)\right] 
\nline
&=& \sum_{l~{\rm even}} \Delta^2_l(k) q_{0l}
\label{eq:cw}
\ear
with,
\bear
q_{0l} = \int_{\mu_{\rm min}}^1 \der \mu~{\cal P}_l(\mu)
\label{eq:qval}
\ear
In the linear (or quasi-linear) models of redshift space distortions, the only values of $l$ that contribute to the sum are $0, 2, 4$ \citep{2012MNRAS.422..926M}. In that case, one can write an explicit expression for the clustering wedge as
\bear
\Delta^2_{0, {\rm win}}(k) &=& \Delta^2_0(k) + \f{1}{2} \mu_{\rm min} \left(1 + \mu_{\rm min}\right) \Delta^2_2(k)
\nline
&+& \f{1}{8} \mu_{\rm min} \left(1 + \mu_{\rm min}\right) \left(7 \mu_{\rm min}^2 - 3\right) \Delta^2_4(k).
\label{eq:deltasq_0_win}
\ear

We found in our earlier paper \citep{2018MNRAS.475..438R} that the recovery of true monopole $\Delta^2_0(k)$, the one corresponding to the full $k$-space average, from the various multipoles inside the window is noisier and less tractable. The model selection is expected to work better if one compares the apparent monopole, i.e., the one obtained in the EoR window. The main goal of this paper is to demonstrate that such a comparison can indeed give unbiased results with error-bars under control. 

\section{MCMC forecasting}

This section outlines various components required for achieving EoR model selection from observations of the 21~cm power spectrum. The estimation of thermal noise and cosmic variance is discussed, which together determine the expected deviations in the power spectrum signal. We give an estimate of these two quantities for the case of SKA1-LOW observations. 

\subsection{Thermal Noise}
\label{sec:thermal}

To estimate the deviation in power spectrum originating the thermal noise we assume a system temperature of \citep{2017isra.book.....T}
\begin{equation}
  T_\mathrm{sys}=60~\mathrm{K}\left(\frac{300~\mathrm{MHz}}{\nu_c}\right)^{2.55},
\end{equation}
where $\nu_c=1420MHz/(1+z)$ is the observation frequency.
The estimation of thermal noise, per Fourier mode measured, is described in \citet{2006ApJ...653..815M, 2012ApJ...753...81P}. The relevant expression is
\begin{equation}
  \Delta_{\mathrm{thermal}}^2(k)\approx X^2Y\frac{k^3}{2\pi^2}\frac{\Omega}{2t}T^2_\mathrm{sys},
  \label{eqn:powerpermode}
\end{equation}
where $k=(k_{\perp}^2+k_{\parallel}^2)^{1/2}$ is magnitude of the Fourier mode, $\Omega$ is field of view of each interferometric element, $t$ is the integration time in seconds for the mode and $X$, $Y$ are cosmology dependent factors which are given as
\begin{equation}
  X\approx 1.9 \frac{h^{-1}\mathrm{cMpc}}{\mathrm{arcmin}} \left(\frac{1+z}{10}\right)^{0.2},
\end{equation}
and
\begin{equation}
  Y\approx 11.5 \frac{h^{-1}\mathrm{cMpc}}{\mathrm{MHz}} \left(\frac{1+z}{10}\right)^{0.5}\left(\frac{\Omega_mh^2}{0.15}\right)^{-0.5}.
\end{equation}
The parallel and perpendicular components of $k$-vector are related to the baseline ${\bf{u}}=$($u,v$) and delay($\eta$) and are given by \citep{2004ApJ...615....7M}
\begin{equation}
{\mathbf k}_{\perp}=\f{2\pi}{D_c(z)}{\mathbf u},
\end{equation}
and
\begin{equation}
k_{\parallel}=\f{2 \pi H_0 \nu_{21}E(z)}{c(1+z)^2}\eta,
\end{equation}
where we have already defined $D_c(z)$ as the comoving distance to redshift $z$ of observation. The quantity $\nu_{21}=1420MHz$ is the rest frequency of 21~cm line and $E(z)\equiv H(z)/H_0$.

Equation~(\ref{eqn:powerpermode}) gives the thermal noise for a mode $k$ that is sampled for time $t$. As the Earth rotates, baseline orientations with respect to the field of view change and thus $\mathbf{k}_{\perp}$ corresponding to a particular baseline will also change. So a given $\mathbf{k}$-mode can be sampled coherently only for a time $t_{\rm {per\mbox{-}mode}}$, the duration after which the baseline starts sampling a different $\mathbf{k}$-mode which can have a different phase. Note that although the modes sampled before and after time $t_{\rm {per\mbox{-}mode}}$, would be close to each other in the ${\mathbf{u}}$-plane, they would be carrying different phases. The estimate of $t_{\rm {per\mbox{-}mode}}$ is based upon the time required for winding of this phase by, say, $\pi$-radians \citep{1999ASPC..180.....T}. Assuming that the measurements of same mode on different days can be added coherently, the total time for sampling the mode coherently can be given as $t=t_{\rm{per\mbox{-}mode}} \cdot t_{\rm{days}}$.
    
Now for doing a comparison with theory, one would be using some sort of binning in $k$-space, and assuming that the binning is small enough, the thermal noise associated with a particular bin would depend on the thermal noise associated with the coherent measurement of the central mode of that bin and the total number of independent coherent measurements, $N_{\rm{per\mbox{-}bin}}$, that fall within that bin. Thus we can write
\be
\Delta_{\mathrm{thermal}}^2({\mathbf{k}})\approx X^2Y\frac{k^3}{2\pi^2}\frac{\Omega}{2~t_{\rm{per\mbox{-}mode}}~t_{\rm{days}}} \f{T^2_\mathrm{sys}}{(N_{\rm{per\mbox{-}bin}})^{1/2}}.
\label{eq:Delta^2_T}
\ee  
Now $N_{\rm{per\mbox{-}bin}}$ is approximately the number of times the central mode can be sampled per day times the number of different samples that belong to the same ${\mathbf k}\equiv (k_{\perp},k_{\parallel})$ bin. The first number is $t_{\rm{per\mbox{-}day}}/t_{\rm{per\mbox{-}mode}}$. The second number will be the number of ${\bf u},\eta$ combinations that fall in the same $(k_{\perp},k_{\parallel})$-bin, we call this quantity $N(k_{\perp},k_{\parallel})$. 

This approach of calculating the thermal noise closely follows the one of \citet{2012ApJ...753...81P} but we make the appropriate changes required for utilizing the sampling in the complete $k_{\perp},k_{\parallel}$ plane. The aim of \citet{2012ApJ...753...81P} was to get an estimate of thermal noise for an experiment like PAPER. Note that PAPER \citep[see for example][]{2015ApJ...809...61A} has predominantly short baselines which means that $k_{\perp}$ is small and much smaller than $k_{\parallel}$ and so $k = \sqrt{k^2_{\parallel}+k^2_{\perp}} \approx k_{\parallel}$. In this scenario, the sampling in the $k_{\perp},k_{\parallel}$ plane appears like a narrow strip with the long edge along the $k_{\parallel}$ direction. As $k \approx k_{\parallel}$, the number of modes available for estimating power spectrum for a particular $k$-value, the $k_{\parallel}$ can be fixed to be $k$ and $k_{\perp}$ can be allowed to be arbitrary. Thus, one can consider slices in $k_{\perp},k_{\parallel}$ plane which have fixed $k_{\parallel}$ value and which extend all the way from $k^{\rm{min}}_{\perp}$ to $k^{\rm{max}}_{\perp}$. 

A similar approach for estimating the thermal noise for different upcoming telescopes was performed in \citet{2016MNRAS.463.2583K}. Our aim is to estimate the noise inside EoR window and compare it to the noise in total $k$-space and in order to do so we would need to find the baseline distribution and compute how many of them lie within the EoR window and how many remain outside. Our final answer for the thermal noise for the entire $k$-space is not very different from the one of \citet{2016MNRAS.463.2583K}. Our way of calculating the noise could also be useful if one wants to combine various regions of $k$-space (e.g., the EoR window, the region between the horizon wedge and the FoV wedge and the region below the FoV wedge) properly. 

The approach of \citet{2012ApJ...753...81P} accounted for the fact that the PAPER field of view is large, about $\Omega_0=0.76 {\rm {sr}}$, and so for the large baselines, the rotation of Earth restricts sampling time per mode $t_{\rm{per\mbox{-}mode}}$ to a fairly short value which is roughly given by
\be
t_{\rm {per\mbox{-}mode}} = t_{20} \left[ \f{\Omega_0}{\Omega} \right]^{\f{1}{2}} \left[ \f{20}{|{\mathbf u}|} \right].
\label{eq:tpm}
\ee
The sampling time $t_{20}$, corresponding to the baseline of length 20 wavelengths, was calculated to be about 13 minutes. One then has to add different samplings of the same baseline on a given day independently instead of coherently. 

As the SKA field of view $\Omega_{\rm FoV} \approx 4.6 \times 10^{-3} {\rm {sr}}$ is relatively smaller\footnote{http://astronomers.skatelescope.org/documents/{\small SKA-TEL-SKO-DD-001-1\_BaselineDesign1.pdf}} (at the frequency of observation corresponding to $z = 9$ and corresponding to a station diameter $d_{\rm{station}} = 35$~m), the sampling time can be very large. But SKA baselines are also larger in lengths (as compared to PAPER baselines) and so these two factors would tend to have opposing effects. We have assumed an observation time of 6 hours per day and the number of days, $t_{\rm{days}}$, to be 120.

As our aim is to find thermal noise for all of $k$-space and not just the EoR window, we would have to compute the $N(k_{\perp},k_{\parallel})$ as a function of the two components of $\mathbf{k}$. We use the following procedure for computing $N(k_{\perp},k_{\parallel})$: First we obtain the actual baseline distribution calculated using antenna co-ordinates given on SKA-Telescope website\footnote{http://astronomers.skatelescope.org/documents/{\small SKA-TEL-SKO-0000422\_02\_SKA1\_LowConfigurationCoordinates.pdf}}. We then assume that the observations are performed with phase centre pointed towards the zenith. This gives a $(u,v)$ distribution \citep{1999ASPC..180.....T} and hence number of modes as a function of $k_{\perp}$ \citep{2004ApJ...615....7M}. For the dependence on $k_{\parallel}$, we assume a bandwidth $B$ of 10,9 and 8 MHz for redshifts of observation 7,8 and 9, respectively. These bandwidths correspond to redshift intervals, $\Delta z$, of 0.45, 0.51 and 0.56 for redshifts 7,8 and 9 respectively. This choice is not strictly essential but it gives a similar comoving volume for the different redshifts and thus allows us to see the degradation in parameter estimation arising from the thermal noise. This choice also gives similar LOS extents ($D_c^{\rm{LOS}} \sim 150 {\rm{Mpc}}$) for the three redshifts. The choice of $\Delta z \approx 0.5$ or $D_c^{\rm{LOS}} \sim 150 {\rm{Mpc}}$ is also advisable as a larger redshift interval would introduce significant amount of light-cone effects \citep{2012MNRAS.424.1877D,2014MNRAS.442.1491D}, while a shorter range would degrade the parameter constraints. We have also assumed a channel width of 62.5~kHz which corresponds to a $k_{\parallel}$ of $\approx 5 {\rm{Mpc}}^{-1}$ which is much larger than the $k$-values relevant for this work.

Once we have the distribution of baselines in the $(k_{\perp},k_{\parallel})$ plane, we simply use concentric circles of $k$-values ($k=(k_{\perp}^2+k_{\parallel}^2)^{1/2}$) to estimate the number of modes that can be observed for a given magnitude-$k$. For the case of EoR window, we simply ignore the modes that fall outside the range, $\mu=[\mu_{\rm {min}},1]$. We then estimate the thermal noise by using equation~(\ref{eq:Delta^2_T}).

\begin{figure}
\includegraphics[width=0.5\textwidth]{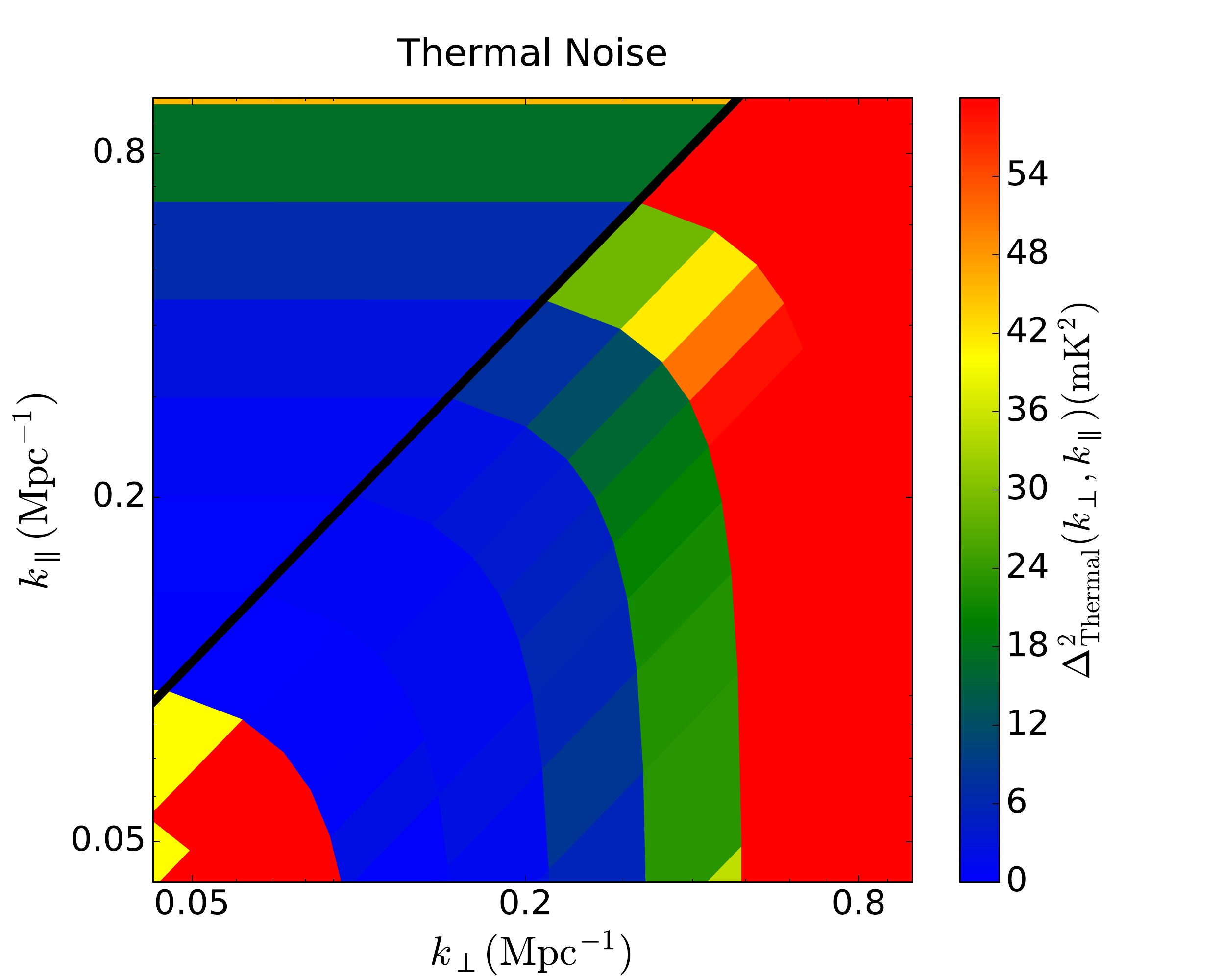} 
\caption{Thermal noise as a function of ($k_{\perp},k_{\parallel}$) for redshift $z=8$ and bandwidth of 9MHz. The black line corresponds to the value of $\mu_{\rm {min}}=0.9$}
\label{fig:ThermalNoisez08} 
\end{figure}

In Figure~\ref{fig:ThermalNoisez08} we estimate the thermal noise for each $(k,\mu)$ bin and plot it assuming a bandwidth of 9MHz centred around the frequency corresponding to $z = 8$. This is done by calculating the number of independent measurements in a given $(k,\mu)$ bin and using equation~(\ref{eq:Delta^2_T}).
  
As seen from the figure, the thermal noise in the last $\mu$-bin, the one closest to $k_{\parallel}$ axis, is the least. This is expected as there are more short baselines as compared to long ones and the spacing along the $k_{\parallel}$ axis is uniform. Also, the longer baselines have shorter integration times as given by equation~(\ref{eq:tpm}) and so samples corresponding to large baselines tend to get added independently (i.e., follow $\approx t_{\rm{tot}}^{-1/2}$ dependence) rather than coherently (i.e. follow $\approx t_{\rm{tot}}^{-1}$ dependence). Note that $t_{\rm{tot}}$ is the total integration time, i.e., 720 hours in our fiducial case. In fact, the thermal noise inside the EoR window is lower by a factor of only $\approx 1.5 - 2$ as compared to the thermal noise for the complete $k$-space result. Thus, even if one adopts foreground avoidance strategy and ignores the samples in the wedge completely, one does not necessarily end up compromising too much on noise characteristics. This observation will be important when we analyse the results of MCMC sampling in the next section. 

\subsection{Cosmic variance}

\begin{figure*}
\includegraphics[width=1.0\textwidth]{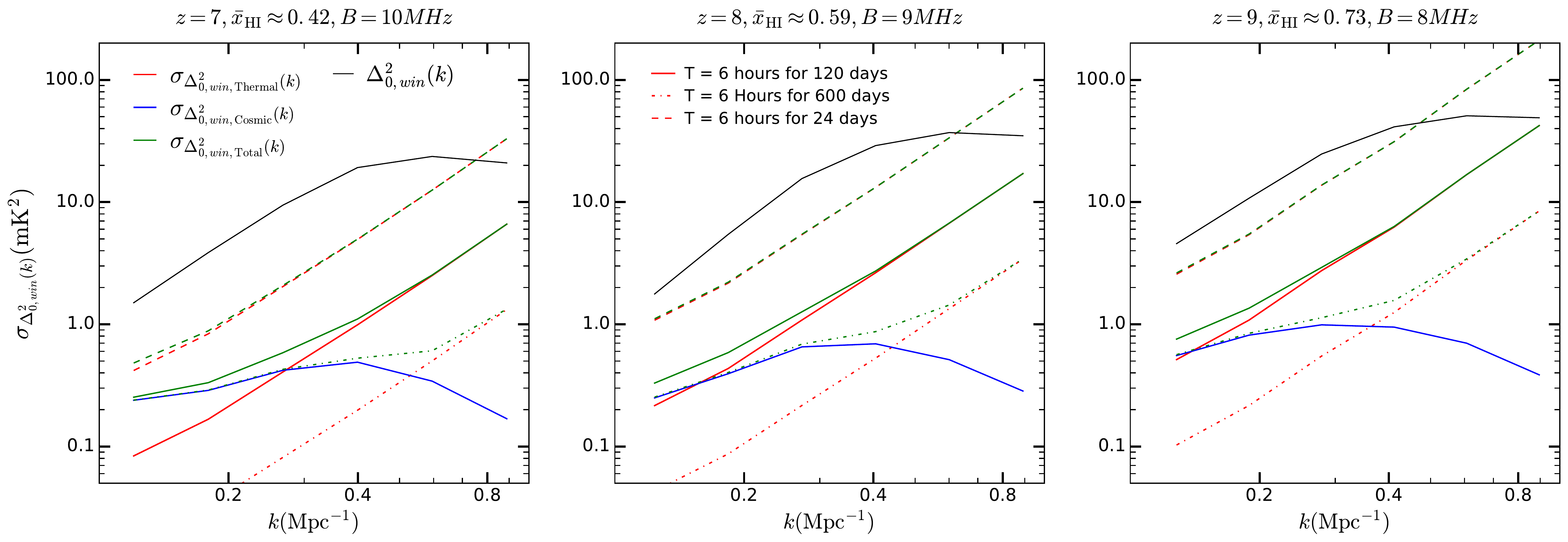} 
\caption{Effect of Integration time on total standard deviation in the EoR window as a function of $k$ for 3 redshifts $z=7,8,9$ with bandwidths of observation 10MHz,9MHz,8MHz.}
\label{fig:ThermalNoiseCompWinAll} 
\end{figure*}

For computing the cosmic variance, we proceed as follows: First, the power spectrum is obtained using the Fourier transformed brightness temperature distribution. The redshift space distortions would render the signal anisotropic but would be preserving the azimuthal symmetry about the LOS axis. Thus we parametrize each ${\mathbf k}$ mode by its magnitude $k$ and the cosine of the angle ($\mu=\cos \theta$) that the mode makes with the LOS direction. After implementing the binning in $(k,\mu)$ space we compute the mean and standard deviation in the power spectrum measurement for each bin. The computation of standard deviation on the power spectrum monopole is then straightforward as each $\mu$ bin contributes equally to a given $k$-mode. As suggested in \citet{2016MNRAS.463.2583K} we then compute the total standard deviation on the monopole by adding the thermal noise and cosmic variance in quadrature: 
\be
\sigma^2_{\Delta^2_0(k)} = \sigma^2_{\Delta^2_0(k),{\rm{Cosmic Variance}}} + \sigma^2_{\Delta^2_0(k),{\rm{Thermal Noise}}}.
\label{eqn:SigmaTot}
\ee 
This relation assumes that the thermal noise and the cosmic variance are completely uncorrelated.

In case of comparing observations in the EoR window, one uses the same expression, with both, the cosmic variance contributions and thermal noise contributions, are now computed using only the modes that are available inside the window. Thus,
\be
\sigma^2_{\Delta^2_{0,win}(k)} = \sigma^2_{\Delta^2_{0,win}(k),{\rm{Cosmic Variance}}} + \sigma^2_{\Delta^2_{0,win}(k),{\rm{Thermal Noise}}}
\label{eqn:SigmaWinTot}
\ee
To account for the fact that our simulation box has a smaller volume as compared to the survey volume\footnote{The survey volume is given by $D_c^{\rm{LOS}} \times A_c$, where $D_c^{\rm{LOS}}$ is the LOS extent corresponding to the bandwidth and $A_c = {\Omega_{\rm{FoV}}} \times D_c(z)^2$ is the comoving area. For the SKA specifications, we find the volume to be $\approx (330 {\rm{Mpc}})^3$, $(360 {\rm{Mpc}})^3$ and $(385 {\rm{Mpc}})^3$ for $z = 7, 8$ and 9 respectively.}, we also compute the square root of the ratio of the two volumes, which is approximately equal to 1.4, 1.8 and 2.2 for the redshifts $z = 7, 8$ and 9 respectively.

\begin{figure*}
\includegraphics[width=1.0\textwidth]{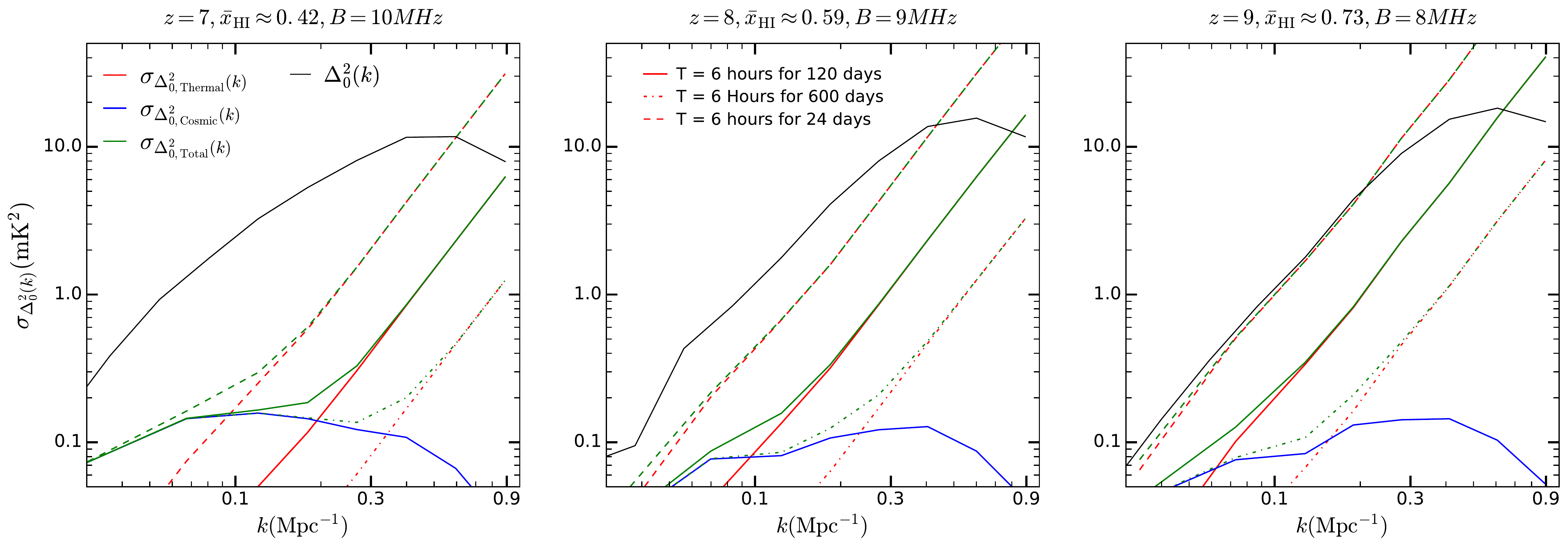} 
\caption{Effect of integration time on total standard deviation as a function of $k$ for 3 redshifts $z=7,8,9$ with bandwidths of observation 10MHz,9MHz,8MHz.}
\label{fig:ThermalNoiseCompAll} 
\end{figure*}

\begin{figure*}
\includegraphics[width=1.0\textwidth]{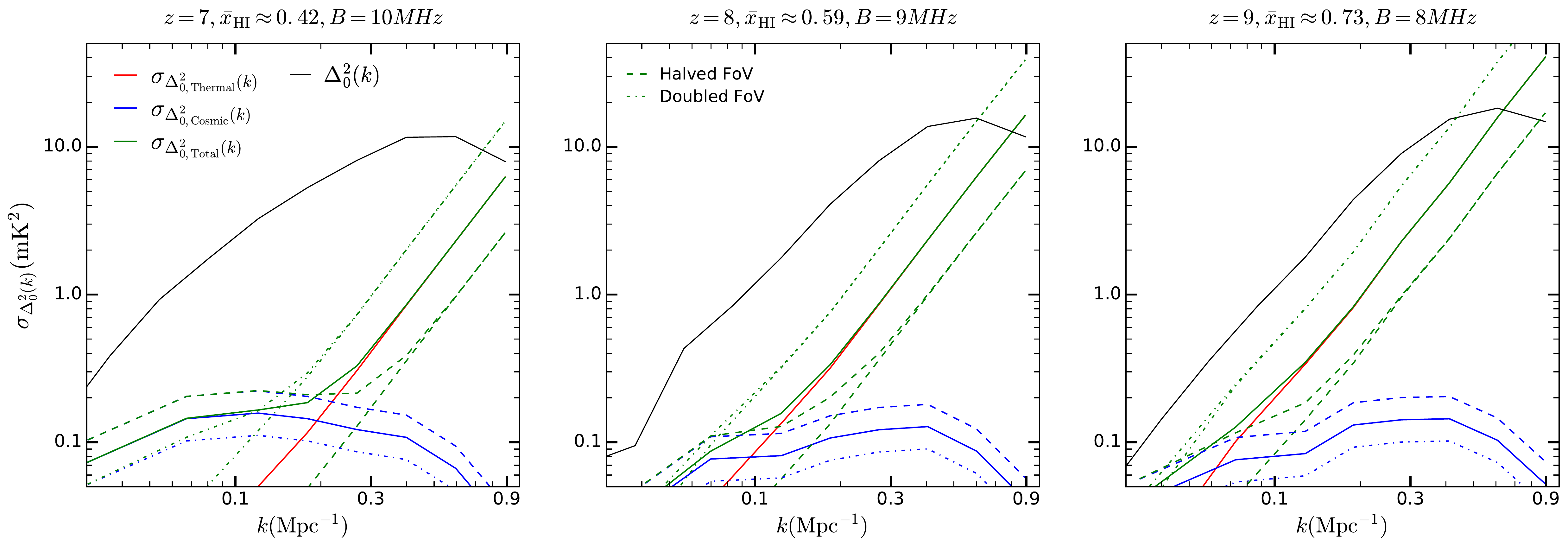} 
\caption{Effect of field of view (FoV) on total standard deviation as a function of $k$ for 3 redshifts $z=7,8,9$ with bandwidths of observation 10MHz,9MHz,8MHz.}
\label{fig:ThermalNoiseCompAllFoV} 
\end{figure*}

To get an understanding of the dependence of total standard deviation on various contributing factors, we plot it for different integration times (Figure~\ref{fig:ThermalNoiseCompAll}) and for different FoV sizes (Figure~\ref{fig:ThermalNoiseCompAllFoV}) for the three redshifts as a function of mode $k$.  As seen from Figures~\ref{fig:ThermalNoiseCompAll} and ~\ref{fig:ThermalNoiseCompAllFoV}, the signal stays above the net standard deviation for most of the cases for $k \lesssim 1.0 {\rm {Mpc}}^{-1}$. It is also seen that the thermal noise is much larger than the cosmic variance for most of the $k$-modes belonging to the range $0.1 {\rm{Mpc}}^{-1} \lesssim k \lesssim 1.0 {\rm {Mpc}}^{-1}$. It is also evident that to make sensitive observations at higher redshifts, say $z=9$,  one should not decrease the integration time significantly or increase the field of view to a much larger value. Both the actions would drive the thermal noise towards the signal and make parameter estimation fairly difficult. However, if one tries to compensate for this by increasing the bandwidth of the observation, the estimation of cosmological signal would tend to degrade. This will happen because at these redshifts the EoR signal is expected to change fairly rapidly (for our fiducial reionization model) and hence if one tries to sample it from a larger redshift range it will tend to get more and more affected by light-cone effects.

In Figure~\ref{fig:ThermalNoiseCompWinAll} we plot the total standard deviation in the EoR window for different integration times for the three redshifts as a function of mode $k$. The lowest two $k$-modes that were shown in the Figure~\ref{fig:ThermalNoiseCompAll} are excluded in this figure as the clustering wedges are not able to give good estimates of power spectrum in the window for these two modes. This happens as there are not sufficient samples available to estimate the quadrupole and hexadecapole accurately at such a low values of $k$. As seen from the figure, the thermal noise is larger in this case, but so is the signal. The separation of the curves for net standard deviation and the power spectrum seem to have gone up maximally for the case of $z=9$. This means that use of only the mean values of power spectrum can possibly even yield better constraints if one uses the EoR window. We would discuss this point further in the next section.

\section{Validation of the model selection procedure using clustering wedges}
\label{sec:valid}

Given that we have the formalisms for estimating the signal and noise in place, we can apply the appropriate statistical analysis tools to validate the model selection using clustering wedges. We use widely applied Fortran code {\sc cosmomc}\footnote{\tt https://cosmologist.info/cosmomc/}\citep{2002PhRvD..66j3511L} for performing MCMC sampling and python package {\sc getdist} \footnote{\tt https://github.com/cmbant/getdist} for analysing the sampled chains and generating the contour plots. As mentioned earlier, we first perform an $N$-body simulation and then smooth the density distributions on a fairly coarse grid to obtain the overdensity field. This can be used to generate a halo catalogue (or equivalently a distribution of $f_{\rm{coll}}$). This is performed for various redshifts and the halo mass distribution in each cell is stored as a function of the minimum halo mass, $M_{\rm {min}}$. The 21~cm brightness temperature distribution for each redshfit can then be generated using the overdensity field and $f_{\rm{coll}}$ field for various values of parameters, $M_{\rm {min}}$ and $N_{\rm {ion}}$. The power spectrum $\Delta^2({\mathbf k})$ of the HI signal and the corresponding Legendre moments can then be calculated for various combinations of parameters as required by {\sc cosmomc}. The standard deviations in the power spectrum values were estimated by methodology described in the previous section. The likelihood $\mathcal{L}$ would be computed through the expression:
\be
- \ln \mathcal{L} = \sum_i \f {\left[\Delta^2_0(k_i)^{\rm {fid}} - \Delta^2_0(k_i)^{\rm {model}}\right]^2}{2\sigma^2_{\Delta^2_0(k_i)}}.
\label{eq:like}
\ee
In case of doing model comparison in the EoR window, both the power spectra and standard deviations were calculated using clustering wedges 
\be
- \ln \mathcal{L} = \sum_i \f {\left[\Delta^2_{0,win}(k_i)^{\rm {fid}} - \Delta^2_{0,win}(k_i)^{\rm {model}}\right]^2}{2\sigma^2_{\Delta^2_{0,win}(k_i)}}. 
\label{eq:likewin}
\ee
For full $k$-space, we perform the comparison using eight $k$-modes (equispaced in $\log k$) ranging from about $0.06 {\rm{Mpc}}^{-1}$ to $0.9 {\rm{Mpc}}^{-1}$ and for the EoR window, we perform the comparison using 6 $k$-modes (again equispaced in $\log k$) ranging from about $0.12{\rm{Mpc}}^{-1}$ to $0.9{\rm{Mpc}}^{-1}$. The very small $k$ modes cannot be included while working in the EoR window as they need accurate computation of higher order multipoles which is not possible for those $k$ values.

The fiducial power spectrum is calculated for a set of input parameters as given in the table~\ref{tab:fidpar}, where we also quote the corresponding neutral hydrogen fractions. As mentioned earlier, the parameter values are chosen so as to match the fiducial reionization history of Figure~\ref{fig:hist}. 
\begin{table}
    \begin{center}
    \caption{Fiducial values of parameters for various redshifts.}
    \label{tab:fidpar}
    \begin{tabular}{lccccc}
      \hline
      Redshift & $\log_{10}M_{\rm {min}}(M_\odot)$  & $N_{\rm {ion}}$ & ${\bar{x}}_{\rm {HI}}$ & B & $\Delta z$ \\ 
      \hline
      z=7 & 8.0 & 6.12 & 0.420 & 10MHz & 0.45 \\
      \hline
      z=8 & 8.0 & 6.22 & 0.595 & 9MHz & 0.51 \\
      \hline
      z=9 & 8.0 & 6.52 & 0.728 & 8MHz & 0.56 \\
      \hline
    \end{tabular}
    \end{center}
\end{table}

\begin{figure}
\centering
\vspace{-0.4cm}
\begin{tabular}{@{}c@{}}      
%\begin{figure}
\includegraphics[width=0.4\textwidth]{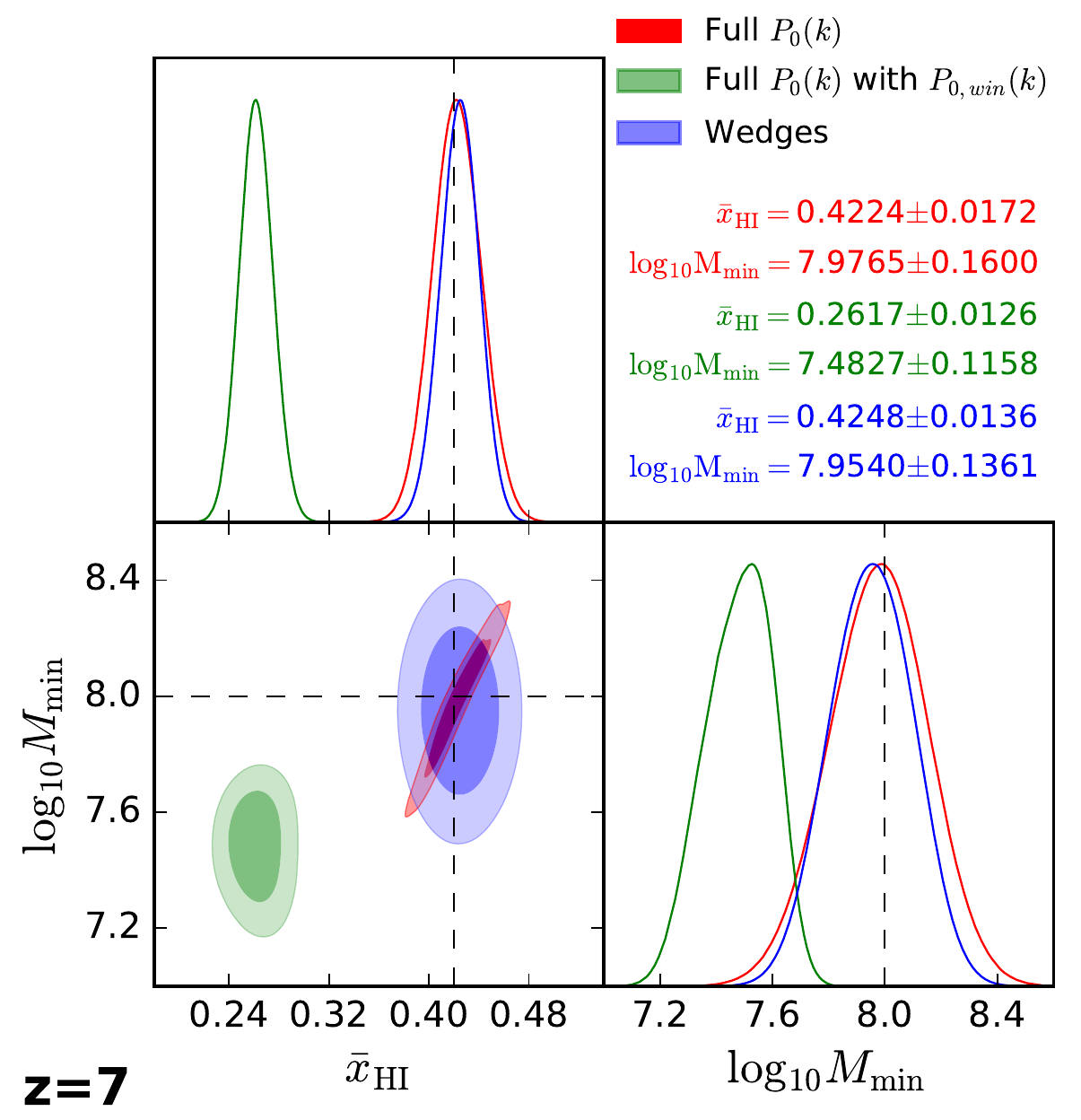}
%\end{figure}
\\
\\
%\begin{figure}
\includegraphics[width=0.4\textwidth]{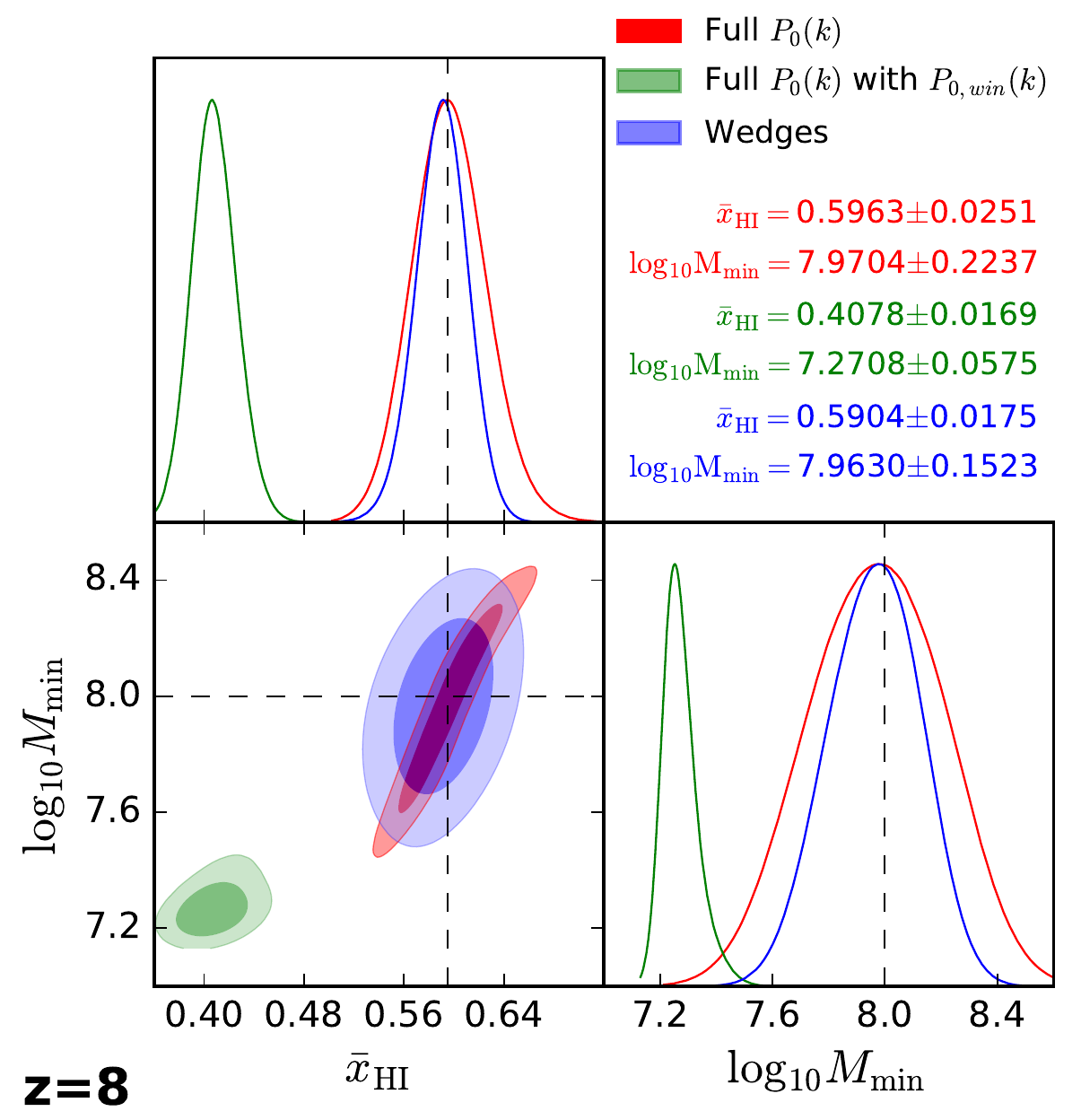}
%\end{figure}
\\
\\
%\begin{figure}
\includegraphics[width=0.4\textwidth]{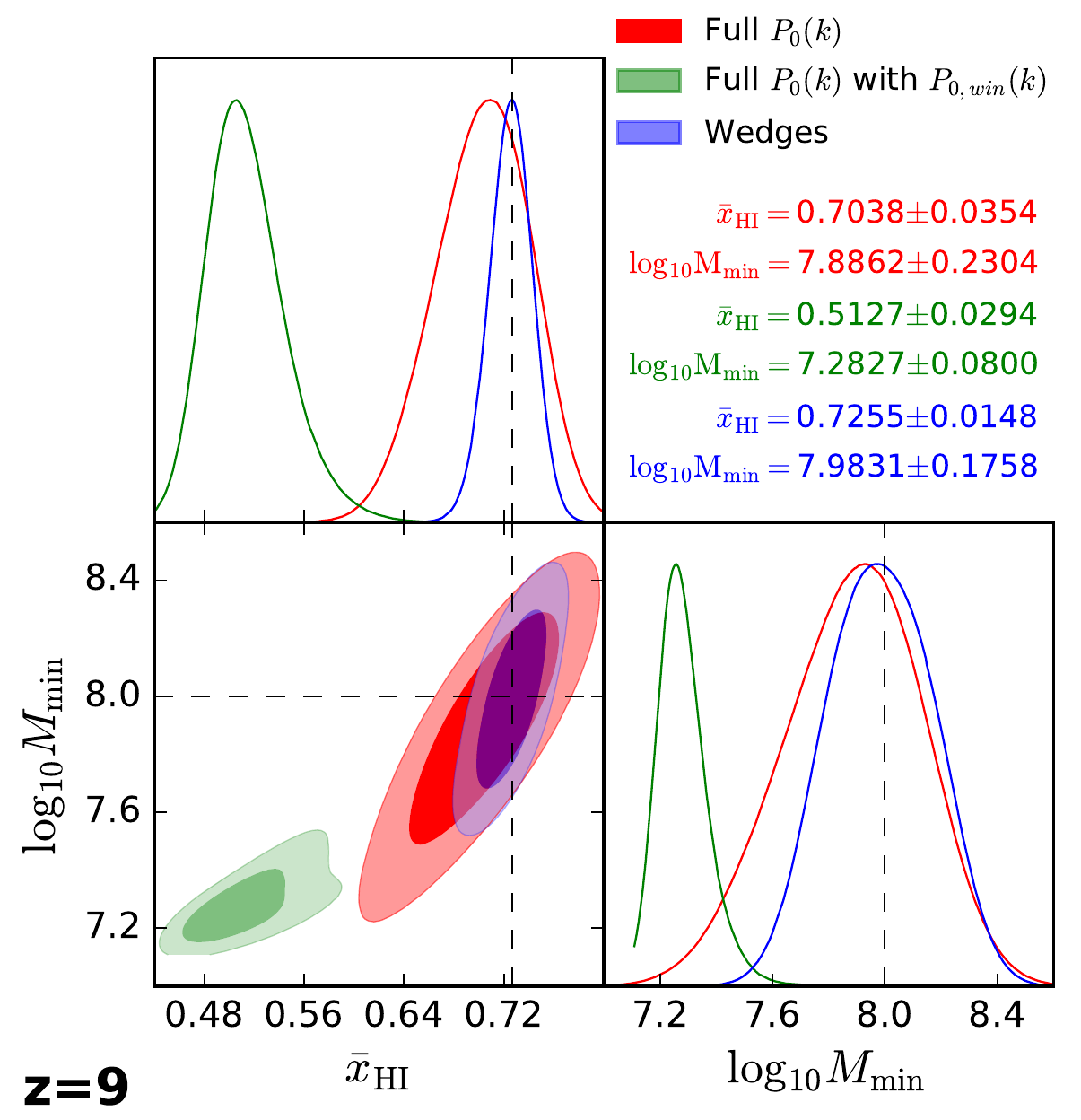}
%\end{figure}
\end{tabular}
\caption{$1\sigma$ and $2\sigma$ parameter constraints arising from the MCMC model selection procedure. The results are given for three different redshifts and for three possible ways of doing the analysis. The contour and parameter estimates are matched in color. Black lines give the fiducial values of the parameters.}
\label{fig:MCMC}
\end{figure}

To demonstrate that the clustering wedges can give unbiased estimates of parameters, we perform the MCMC sampling using the true power spectrum monopole as well as using clustering wedges and then compare the results. We treat $M_{\rm min}$ and $\bar{x}_{\rm HI}$ as the free parameters. These two parameters are equivalent to the parameters $T_{\rm vir}$ (the minimum virial temperature of haloes that can form stars) and $N_{\rm ion}$ used by \citet{2015MNRAS.449.4246G}. The three different panels in the Figure~\ref{fig:MCMC} correspond to the MCMC results for three different redshifts mentioned in the table. Each panel of the figure has 3 sets of contours: (i)~The red contours correspond to the case of comparing full $k$-space power spectra for parameter models with full $k$-space power spectrum of the fiducial model, which would be the case where one has access to the full $k$-space while doing the observations. (ii)~The green contours correspond to the comparison of full $k$-space power spectra for parameter models with the partial $k$-space power spectrum of the fiducial model, which is equivalent to the case where the data is available only in the EoR window but the theoretical calculations are performed in the full $k$-space. (iii)~The blue contours correspond to the case of comparing the clustering wedges of parameter models with the partial $k$-space power spectrum of the fiducial model, which is simply equivalent to working only in the EoR window. The contours are shown for $1\mbox{-}\sigma$ and $2\mbox{-}\sigma$ confidence levels.

As we can see in Figure~\ref{fig:MCMC}, the full $k$-space analysis, as expected, recovers the input fiducial values with reasonably high accuracy (red contours). However, as can be seen from the contours, the predictions are also degenerate. The origin of degeneracy is explained towards the end of this section.

However, the comparison of complete $k$-space power spectrum with partial one (green contours) predicts parameter values which are off from the input values by several standard deviations. The $\bar{x}_{\rm{HI}}$ mean values are off by about 12, 11 and 7 $~\sigma$ while the $M_{\rm{min}}$ mean values are off by about 4.5, 12 and 9 $~\sigma$ from the fiducial values for the top, middle and bottom panels respectively. This is simply the wedge bias manifested in terms of the EoR parameters.

The analysis using clustering wedges (blue contours) removes the wedge bias and the predictions are consistent with the input values. The parameter constraints obtained using clustering wedges are poorer for $z=7$ and $z=8$ case when compared to the full $k$-space, as is evident from the area of the contours. This is because the thermal noise is larger when working only in the EoR without any significant increment in the power spectrum amplitudes. The clustering wedges, although, do seem to reduce the extent of degeneracy that seems to be there between the two parameters and hence the errors on the individual parameters are similar to that in the case of the full $k$-space.

One can see from the bottom panel of Figure~\ref{fig:MCMC} that the clustering wedges constraints are better  compared to the complete $k$-space result for redshift $z=9$. The reason behind this is that, at high redshifts, the cross-correlation of baryonic and HI density fluctuations is positive which leads to a larger quadrupole of the 21~cm fluctuations. This, in turn, leads the power spectrum measured in the window to be boosted up. One can see from Figures~\ref{fig:ThermalNoiseCompAll} and \ref{fig:ThermalNoiseCompWinAll} that the ratio of the power spectrum to its net standard deviation at $z = 9$ is larger for the EoR window than that for the full $k$-space.

\begin{figure*}
\centering
\begin{tabular}{@{}lccr@{}}      
\includegraphics[width=0.4\textwidth]{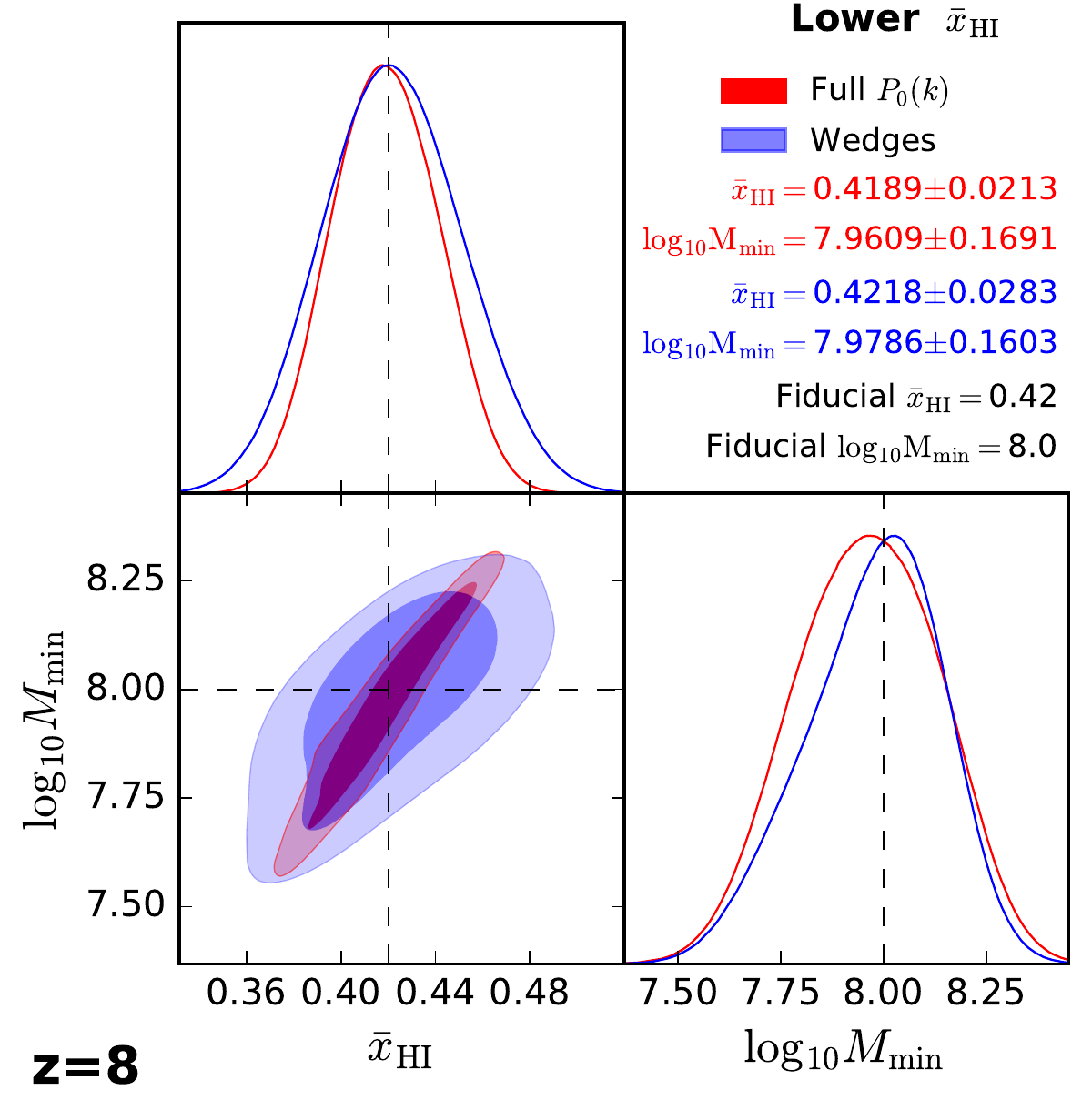}&
&
&
\includegraphics[width=0.4\textwidth]{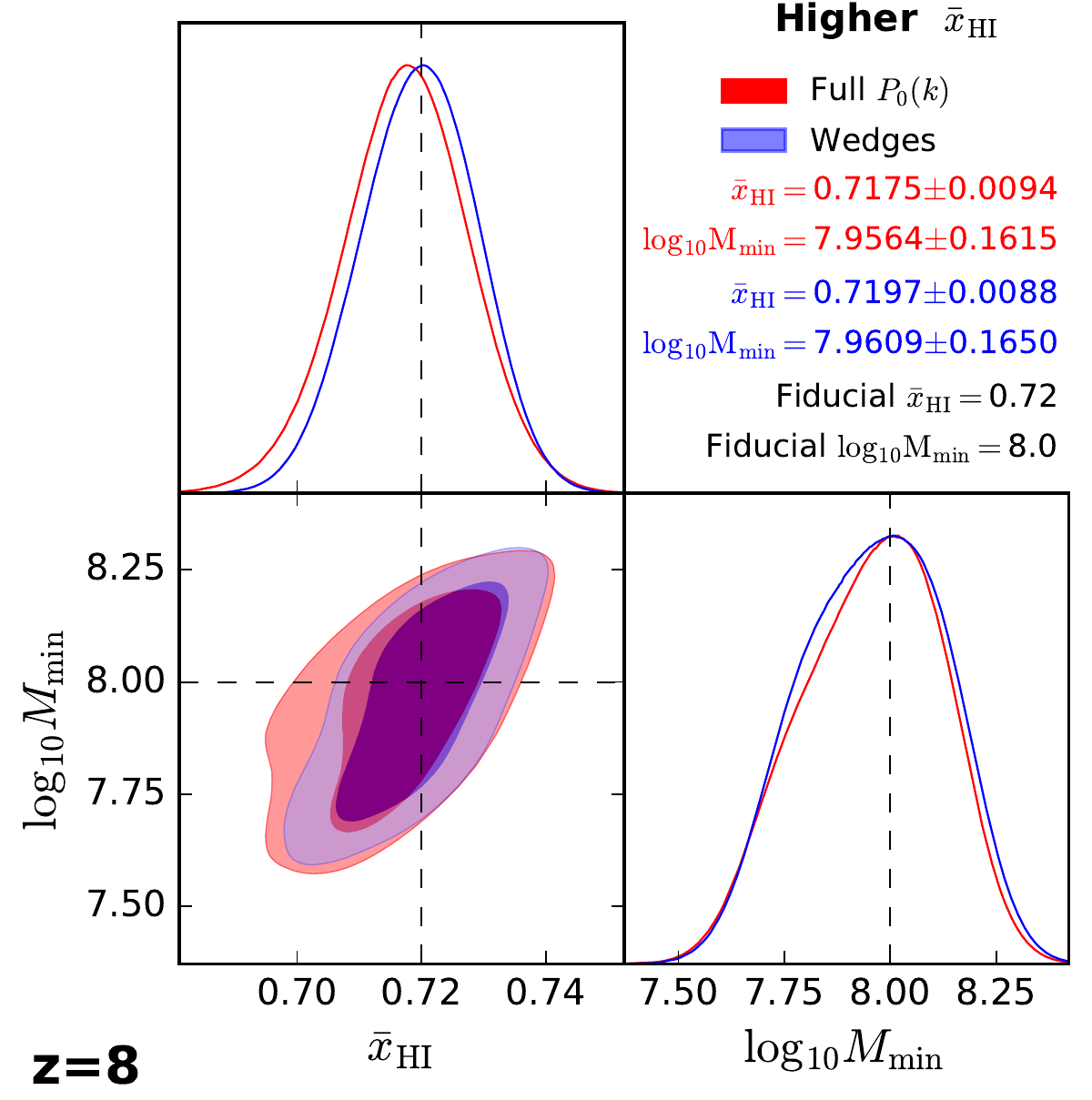}
\end{tabular}
\caption{MCMC results for $z=8$ done with a lower and higher value of mean neural hydrogen fraction.}
\label{fig:MCMC2}
\end{figure*}

To confirm our point of getting comparable constraints using clustering wedges due to the enhancement of the signal, we do more analysis, this time for the same redshift but for a different mean neutral hydrogen fraction. We choose the middle redshift box ($z=8$) and change the fiducial model to be one having higher and lower mean neutral hydrogen fractions. These fractions were chosen to match to the ones corresponding to the other two analysed redshifts. The results are as depicted in Fig~\ref{fig:MCMC2}. As seen from the left plots of Figure~\ref{fig:MCMC2}, the parameter constraints degrade (as seen from the significant increase in the contour areas) with the use of clustering wedges and the behaviour is similar to the one corresponding to the lower redshift box. On the other hand, for the higher mean neutral hydrogen fraction, the constraints get better (as seen from the decrease in the contour areas) and behaviour becomes similar to the one for higher redshift box.

We end this section by commenting on the degeneracies in the parameter constraints in our analysis. We find that the parameters constraints are highly degenerate when using the true monopole in the full $k$-space while the degeneracy is less when we use the clustering wedges in the EoR window. For the scales of interest, the monopole of the power spectrum is dominated by the contribution from fluctuations in the ionization field. For a given $M_{\rm min}$, these fluctuations decrease with increasing $\bar{x}_{\rm HI}$ for $\bar{x}_{\rm HI} \gtrsim 0.4$. On the other hand, for a fixed $\bar{x}_{\rm HI}$, the ionization fluctuations increase with increasing $M_{\rm min}$ as the field is sourced by massive and rare haloes. This leads to a positive correlation between $M_{\rm min}$ and $\bar{x}_{\rm HI}$ in our analysis. A similar degenerate behaviour for entire $k$-space analysis is also seen in the earlier works of \citet{2015MNRAS.449.4246G} and \citet{2014ApJ...782...66P}. The details of the reionization model used there is somewhat different from ours, however, their model parameters are similar to the ones used in this paper.

Restricting the analysis to the EoR window effectively mixes the higher order multipoles to the monopole. In fact, a quadrupole contribution $q_{02} \Delta^2_2(k)$ is added to the monopole term along with the hexadecapole contribution $q_{04} \Delta^2_4(k)$. The hexadecapole term is mostly determined by the underlying matter fluctuations which is independent of the EoR parameters, hence it does not affect the parameter degeneracies. The quadrupole term, on the other hand, is dominated by the cross-correlation between the HI and baryonic density fields. In the inside-out models of reionization where high density regions are ionized first, the cross-correlation decreases with decreasing $\bar{x}_{\rm HI}$. For a fixed $\bar{x}_{\rm HI}$, increasing $M_{\rm min}$ leads to ionization field characterized by larger bubbles and hence the correlation between the HI and baryonic field decreases. Thus the nature of degeneracy between the two parameters is opposite in the case of the quadrupole to that in the case of the monopole. These two opposite effects lead to degradation in the degenerate behaviour between the parameters. 

We thus see that there is a generic degradation in the parameter constraints when working only in the EoR window as compared to the full $k$-space, which is somewhat compensated by the breaking of degeneracies in the parameter constraints.

\section{Comparing Foreground avoidance with Foreground modelling}
\label{sec:compFGFM}

The discussion in this paper has so far been concentrated on testing whether clustering wedges can yield faithful and reasonable parameter constraints. It is interesting to examine how the foreground subtraction/removal method compares with the avoidance technique. A detailed comparison of the two approaches can also be found in \citet{2014arXiv1408.4695C} where they perform an advanced Generalized Morphological Component Analysis (GMCA). In this approach one tries to utilize the smooth spectral nature of the foregrounds to find a suitable basis so as to fit them with very few components. They find that the foreground removal works better for the case of LOFAR, however, for the case of SKA, the foreground removal seems to underestimate the signal component by overfitting for the foreground part.   %Another approach to the problem of foreground removal is to use a Bayesian framework to jointly constrain the foreground and EoR power spectra \citet{2016MNRAS.462.3069S}. They have modelled their analysis for the experiment HERA \footnote{http://reionization.org/} and found that the intrinsic instrumental power spectra are recovered well.

In this section, we will briefly try to compare the approach of foreground avoidance which relies on using clustering wedges with the approach of foreground modelling where one has access to entire $k$-space. This comparison can only be approximate as the deviation in power spectrum due to foreground residuals cannot be said to have been perfectly understood yet. This deviation may not only depend on the telescope properties, but also on the understanding and modelling of the processes that give rise to the various types of foregrounds. Here we adopt estimates of the residual power spectrum for SKA1 observations give by \citet{2015MNRAS.447.1973B} where they use the correlated component analysis (CCA) method for estimating the foreground contamination and filter out the 21~cm signal from the simulated temperature maps. The level of foreground residuals obtained using their method depend on the actual model used, however, we find that the residuals are about 10\% of the cosmological signal for $z \approx 8$. In absence of a full understanding of how to subtract out the foregrounds, we use this approximate value as an estimate of foregrounds in our analysis. For model comparison we thus assume additional contribution coming from foregrounds as 
\be
\sigma_{\Delta^2_0(k),{\rm {Foregrounds}}} \approx 0.1 \times \Delta^2_0(k)
\ee 
and add this quantity to earlier expression of standard deviation
\be
\sigma^2_{\Delta^2_0(k)} = \sigma^2_{\Delta^2_0(k),{\rm{Cosmic Variance}}} + \sigma^2_{\Delta^2_0(k),{\rm{Thermal Noise}}} + \sigma^2_{\Delta^2_0(k),{\rm {Foregrounds}}}.
\ee

\begin{figure}
\includegraphics[width=0.4\textwidth]{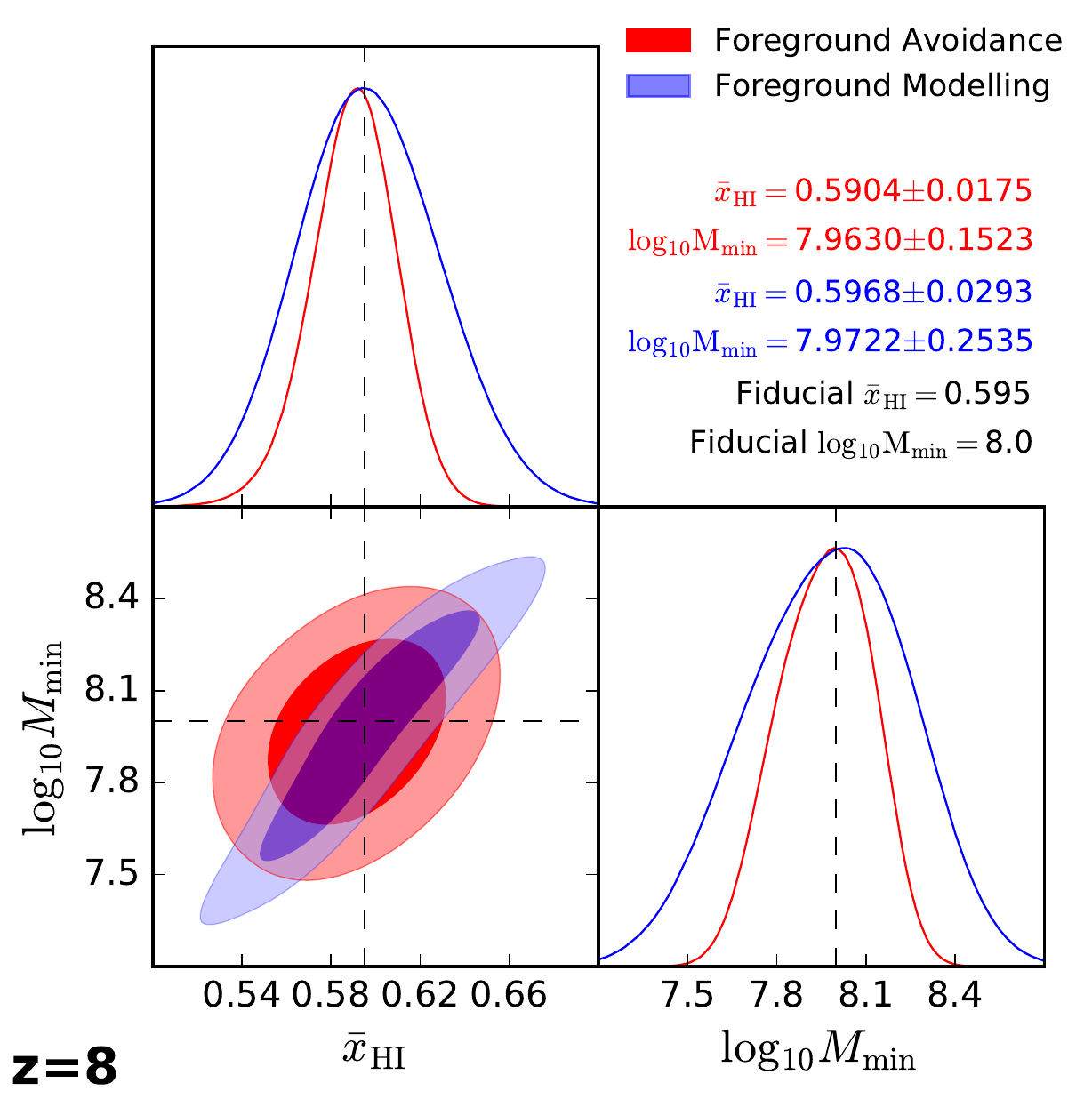}
\caption{MCMC done for the $z=8$ box, with foreground modelling approach. Also shown are the results obtained using clustering wedges (Foreground Avoidance).}
\label{fig:QuadInc}
\end{figure}

The results of model selection using this likelihood are as depicted in Figure~\ref{fig:QuadInc}. 
As seen from Figure~\ref{fig:QuadInc}, the areas of the parameter contours are smaller for the foreground modelling case than the avoidance. However, the parameters are highly degenerate in the foreground modelling case which is expected from our earlier discussions. Because of this degeneracy, the constraints on the individual parameters turn out to be worse when we attempt to model and subtract the foreground. Interestingly, the constraints obtained using the foreground modelling approach are comparable to the ones obtained by \citet{2015MNRAS.449.4246G}, where they get $\sigma_{\bar{x}_{\rm{HI}}}$ to be of about 0.01 and 0.04 (for redshift 8) for zero and 25\%  EoR modelling uncertainties scenarios respectively with a total integration time of 1000 Hours.

We should reiterate that the model we use for foreground residuals is probably too simplistic, and it is possible that the errors estimated by us could be reduced with better analysis. One can still conclude that using the clustering wedges help breaking the degeneracy in the two parameters and thus obtain error-bars that are competitive. One possible approach to further improve the constraints could be to combine the two approaches which we defer for the future.

\section{Summary and discussion}

The presence of very large astrophysical foregrounds makes the extraction of 21~cm signal from the Epoch of Reionization a very challenging endeavour. An appropriate interpretation of the signal, achieved through foreground avoidance approach, on the other hand, does require a careful analysis using Legendre polynomials. A framework of model selection that does not get biased due to the presence of foreground wedge structure is essential for developing a faithful understanding of various astrophysical parameters. In this paper, we build upon our previous understanding \citep{2018MNRAS.475..438R} to examine whether the use of clustering wedges can provide reliable and unbiased parameter constraints in the EoR. We make use of semi-numerical simulations and couple them with MCMC-based statistical methods to obtain the parameter constraints. We also model the thermal noise carefully by taking into consideration the proposed baseline distribution of the SKA1-Low. We concentrate on the spherically averaged power spectrum, or the monopole, while computing the likelihood for the MCMC analysis.

We find that the constraints obtained on individual EoR parameters $\bar{x}_{\rm HI}$ and $M_{\rm min}$ using the clustering wedges are comparable to the constraints obtained using the knowledge of all the Fourier modes. The main reason for getting comparable results (in spite to having access to a limited number of $k$-modes) is the anisotropic nature of thermal noise in the $k_{\perp} - k_{\parallel}$ space. Essentially the excess in the number of modes with short baselines, which also possess larger sampling times, renders the determination of corresponding power spectrum more robust. 

The clustering wedges also help in breaking the degeneracy between $\bar{x}_{\rm HI}$ and $M_{\rm min}$ which is present in the case of using all $k$-modes. This is because the partial $k$-space coverage leads to mixing of the higher order multipoles (i.e., the quadrupole and the hexadecapole) with the monopole term.  The dependence of the quadrupole on the $\bar{x}_{\rm HI}$, in fact, is opposite to that for the monopole, which helps in lifting the degeneracy.

There are various possible extensions of our work. Note that because of the anisotropic nature of both signal and thermal noise, the parameter estimation procedure for the real data should be done more carefully. The optimal sampling which can yield the best possible constraints needs to be designed in such a way that, while giving a better signal to noise ratio, no compromise is made on obtaining the true value of the signal. A possible way to address this issue could be to use the combined knowledge of the first three even multipoles and use a weighting scheme that preserves evaluation of all them simultaneously. We plan to take up this question in the future.

We have also assumed in this paper that the foregrounds are completely absent in the EoR window, which is probably a oversimplification \citep{2013ApJ...768L..36P}. It would be interesting to include these complications into our analysis and see if the parameter degeneracies get modified.

\section*{Acknowledgements}
The simulations used in the paper were performed
on the IBM cluster hosted by the National Centre for Radio
Astrophysics, Pune, India.

\bibliographystyle{mnras}
\bibliography{main}

\label{lastpage}
\end{document}